%% file: preferential-attachment.tex
\documentclass{sigchi}

\usepackage[utf8]{inputenc}
\usepackage{balance}  % to better equalize the last page
\usepackage{graphicx}
\usepackage{grffile}
\usepackage{times}    % comment if you want LaTeX's default font
\usepackage{url}      % llt: nicely formatted URLs
\usepackage{subfigure}
\usepackage{booktabs}
\usepackage{color}

\toappear{
Permission to make digital or hard copies of all or part of this
work for personal or classroom use is granted without fee provided that 
copies are not made or distributed for profit or commercial advantage and
that copies bear this notice and the full citation on the first page. To
copy otherwise, or republish, to post on servers or to redistribute to 
lists, requires prior specific permission and/or a fee.\\
{\confname{WebSci'13}}, May 2--4, 2013, Paris, France.\\
Copyright 2013 ACM 978-1-4503-1015-4/12/05...\$10.00.
}

\makeatletter
\def\url@leostyle{%
  \@ifundefined{selectfont}{\def\UrlFont{\sf}}{\def\UrlFont{\small\bf\ttfamily}}}
\makeatother
\def\pprw{8.5in}
\def\pprh{11in}

\usepackage{hyperref}
\hypersetup{
pdftitle={
  Preferential Attachment in Online Networks: Measurement and Explanations
},
pdfauthor={Jérôme Kunegis and Marcel Blattner},
pdfkeywords={Network analysis, preferential attachment, web science},
bookmarksnumbered,
pdfstartview={FitH},
colorlinks,
citecolor=black,
filecolor=black,
linkcolor=black,
urlcolor=black,
breaklinks=true,
}

\newcommand{\wOne}{1.00\columnwidth}
\newcommand{\wTwo}{0.48\columnwidth}
\newcommand{\wFullThree}{0.65\columnwidth}

\begin{document}

\title{
  Preferential Attachment in Online Networks: \\ Measurement and Explanations
}

\numberofauthors{3}
\author{
  \alignauthor Jérôme Kunegis\\
    \affaddr{University of Koblenz--Landau}\\
    \email{kunegis@uni-koblenz.de}
  \alignauthor Marcel Blattner\\
    \affaddr{\scalebox{0.9}{Laboratory for Web Science, FFHS}}\\
    \email{marcel.blattner@ffhs.ch}
  \alignauthor Christine Moser\\
    \affaddr{VU University Amsterdam}\\
    \email{c.moser@vu.nl}
}

\teaser{
  \centering
  \includegraphics[width=1\textwidth]{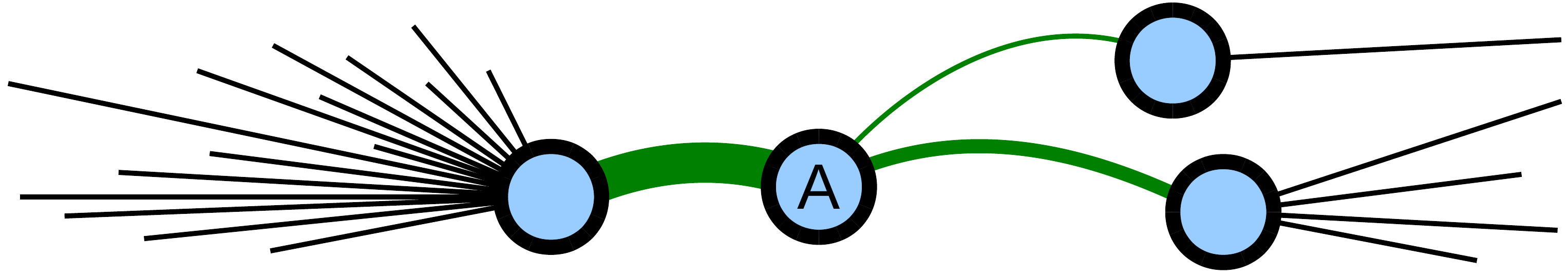}
  \caption{
    Schematic representation of the preferential attachment process:
    The probability 
    that a tie appears between node A and another node is a function of
    the number of ties of the other node. In this example,
    probabilities of new ties (in green) are indicated by line width. 
    In this paper, we measure this relationship between the degree and the
    tie creation probability, modeling it as a power with an exponent
    whose values we explain by the processes underlying the network. 
  }
  \label{fig:teaser}
}

\maketitle

\begin{abstract}
We perform an empirical study of the preferential attachment phenomenon
in temporal networks and show that on the Web, networks follow a
nonlinear preferential attachment model in which the exponent depends on
the type of network considered.  The classical preferential attachment
model for networks by Barabási and Albert (1999) assumes a linear
relationship between the number of neighbors of a node in a network and
the probability of attachment. Although this assumption is widely made
in Web Science and related fields, the underlying linearity is rarely
measured. To fill this gap, this paper performs an empirical
longitudinal (time-based) study on forty-seven diverse Web network
datasets from seven network categories and including directed,
undirected and bipartite networks. We show that contrary to the usual
assumption, preferential attachment is nonlinear in the networks under
consideration.  Furthermore, we observe that the deviation from
linearity is dependent on the type of network, giving sublinear
attachment in certain types of networks, and superlinear attachment in
others. Thus, we introduce the preferential attachment exponent $\beta$
as a novel numerical network measure that can be used to discriminate
different types of networks.  We propose explanations for the behavior
of that network measure, based on the mechanisms that underly the growth
of the network in question.
\end{abstract}




\section{Introduction}
The term \emph{preferential attachment} refers to the observation that
in networks that grow over time, the probability that an edge is added
to a node with $d$ neighbors is proportional to $d$.  This linear
relationship lies at the heart of Barabási and Albert's
\emph{scale-free} network model \cite{b439}, and has been used in a vast
number of subsequent work to model networks, online and offline. The
scale-free network model results in a distribution of degrees, i.e.,
number of neighbors of individual nodes, that follow a power law with
negative exponent. In other words, the number of nodes with $d$ is
proportional to $d^{-\gamma}$ in these networks, for a constant
$\gamma>1$.  While a large amount of work has been done to verify
empirically the
validity of such \emph{power laws} of the degree distribution,
relatively little work has investigated whether the initial assumption
of linear preferential attachment is valid.  The only such study known
to the authors is that of Jeong, Néda and Barabási \cite{b767}, which
observes a preferential attachment function that is a power of the
degree with an exponent in the range $[0.80, 1.05]$. However, that study
investigates only four network datasets representing only a small subset of
network types encountered on the Web, and does not explain why the
networks have a specific value of the exponent,
which is crucial to better understand the dynamics
and social processes that underlie preferential attachment, and thus
the behavior of online networks in general.  
Due to the availability in
recent years of a large number of independent network datasets covering
diverse aspects of the World Wide Web, we are able  to
study the behavior of forty-seven network datasets, and can interpret the
observed values of the preferential attachment exponent in terms of the
social processes underlying the individual networks. 

The contributions of this paper are:
\begin{itemize}
  \item We provide a method for measuring the
    preferential attachment 
    exponent $\beta$ empirically in networks for which temporal
    information is known.
  \item We perform an extensive and systematic study of preferential
    attachment in forty-seven online networks from seven different
    network categories.
  \item We give interpretations for the observed
    values, showing that six out of the seven network categories display
    preferential attachment behavior with values of the preferential
    attachment exponent consistently above or below the value
    $\beta=1$. 
\item We interpret these findings in terms of social processes and
  explain why network categories feature such consistent behavior. 
\end{itemize}

The remainder of the paper is structured as follows. 
In Section \emph{Related Work}, we give a detailed review of
  growth models leading to power law-like distributions, both for
  networks and non-network datasets.  
In Section \emph{Method}, we introduce our method for
  empirically measuring the preferential attachment exponent of a given
  temporal network dataset. 
In Section \emph{Experiments}, we apply our method to a
  collection of forty-seven temporal network datasets. 
Finally, in Section \emph{Explanations}, we give explanations for the
  observed behavior. 

\section{Related Work}

The concept of preferential attachment has received ample attention in
science, in particular network science \cite{b774}. Accompanying the rise of
the World Wide Web, a new generation of studies was published. Aside
from renewed interest in explanatory models (starting with Barabási and
Albert's seminal \emph{Science} paper in 1999 \cite{b439}, see also next
section), social 
scientists started to use the concept in order to explain social
processes. Preferential attachment is generally understood as a
mechanism where newly arriving nodes have a tendency to connect with
already well-connected nodes \cite{b771,b779,b774}.  

Preferential attachment has been used to explain observations in a variety of
networks. For example, Lemarchand \cite{b778} and Wagner and Leydesdorff
\cite{b781} 
explain evolving co-authorship networks based on preferential
attachment. Similarly, Barabási and colleagues \cite{b771} investigate
collaboration networks in science and find that preferential attachment
acts as a governance mechanism in the evolution of these networks. 
Also, Hanaki and colleagues \cite{b777} find that collaboration networks in the IT
industry are significantly related to preferential attachment. Gay and
Dousset \cite{b776} investigated the biotechnology industry, where new firms
attach preferentially to older and ``fitter'', i.e., more successful,
firms. A still emerging field of research is that of online networks, 
continuing the research stream started by Barabási and Albert \cite{b439} more
than a decade ago. For example, Tremayne and colleagues \cite{b780}
investigated preferential attachment in the war blog network, where
links from other blogs and reporting posts were significant
predictors. Most recently, Faraj and Johnson \cite{b775} have investigated open
source software networks and found, surprisingly, a tendency away from
preferential attachment.  

\subsection{Power Laws and Related Distributions}
\label{sec:degree-distribution}

Power laws and related distributions such as the lognormal and Simon--Yule
distributions can 
be understood as the result of generative processes and are observed
in many different areas, e.g., physics, biology, 
astronomy and economics. 
Kapteyn \cite{b762} and Gibrat \cite{b753} are recognized as two of the
first scientists connecting generative processes with lognormal 
distributions, in 1916 and 1933 respectively. 
Champernowne \cite{pre1} showed that a small change in the lognormal
generative process results in a generative process with a power law
distribution.
Yule \cite{b752} explained the observed power law distribution of 
species among genera of plants. 
A clean explanation how preferential attachment leads to a power law
distribution was given by Simon \cite{b754}. 
Zipf \cite{b759} found that word frequencies follow a power law distribution
and Lotka \cite{b760} showed the number of written articles by authors
follows a power law distribution as well.

Recent work on power law distributions is focused on
graph and network structure, e.g., the World Wide Web,
the Internet, collaboration networks and others
in connection with preferential attachment mechanisms \cite{b59,b439,b116}.
Finally, Newman \cite{b462,b408} outlined the challenges in measuring
power law exponents in real data.

%% \paragraph{What Preferential Attachment Is Not}
%% %% do we really need this paragraph? 
%% Preferential Attachment is a statement of the relative growth of
%% multiple node, i.e., a node with double the degree will grow at double
%% the rate.  Preferential attachment does not predict how fast growth will
%% be, i.e., how many nodes will be expected after a time $t$.  This is for
%% instance plotted in~\cite{b751}. 

\subsection{Nonlinear Preferential Attachment}
Traditionally, the preferential attachment function is assumed to be linear,
i.e., directly proportional to the node's degree.
A natural generalization is to an arbitrary function of the
node degree, in particular to sub- and superlinear functions
\cite{b764,b756,b766,b770}. 

Superlinear preferential attachment functions in trees are investigated
in \cite{b765}. The asymptotic distribution of degrees has the
probability $P=1$ for one vertex (the perpetual hub) and the probability
$P=0$ for all other vertices. 
Physically, this is a gelation-like phenomenon. 
In \cite{b769} it is shown that a moderate superlinear preferential attachment
function leads to a degenerate degree distribution in the
thermodynamic limit, where one node receives almost all edges.
However, in a wide range of the \emph{pre-asymptotic} regime the
degree distribution follows a power law distribution. 

In the sublinear case of a preferential attachment function that is a
power with an exponent between zero and one, a stretched exponential
degree distribution is the result \cite{b764}. 

\subsection{Measuring Preferential Attachment}
In most network studies, only a static snapshot of a network is
available, and thus is is not possible to verify empirically whether
preferential attachment takes place.  Instead, most papers study the
degree distribution of a network, and interpret its specific forms as
evidence for preferential attachment. All references cited in the
section \emph{Power Laws and Related Distributions} fall
into this category.  

The preferential attachment exponent itself is measured for several
networks in 
\cite{b767}, where all exponents were found in the range $[0.8,1.0]$. 
Newman \cite{b768} investigated a scientific collaboration network, 
finding a linear preferential attachment up to a degree $\approx 500$, 
and a sublinear preferential attachment beyond that value.
These investigations show that a linear preferential attachment function
is rarely observed in real world data, and that sub- and superlinear functions
play a pivotal role in governing the network growing process
attributed to the preferential attachment mechanism.

%This is the reference that comes closest to our paper. 
%(What we do differently than that
%paper:  we interpret the preferential attachment exponent in terms of
%the processes on the Web.) 

%Empirical evidence for preferential attachment has been found in open
%source software 

%projects~\cite{b751}, based on the Debian community.  The number of new
%inlinks is proportional to the indegree, with a standard deviation
%proportional to the square root of the indegree. 
%This reference states (in slide 26):
%\begin{itemize}
%\item average growth rate in function of old degree:  $R(\Delta t) =
%  \langle\Delta k/k \rangle$  
%\item  standard deviation of growth rate:  $S(\Delta t) = \langle[\Delta k/k]^2\rangle^{1/2}$
%\item prediction:  $R(\Delta t) = \mu \Delta t, S(\Delta t) = \sigma sqrt(\Delta t)$
%\end{itemize}

\section{Method}

The concept of preferential attachment has slightly different
interpretations depending on the type of networks considered. In order
to distinguish these different cases, we 
classify the networks available to us into seven categories, depending
on the underlying 
entities and relationships represented by vertices and edges.

Although some networks are genuinely undirected, such as a friendship
network, many networks represent asymmetric relationships 
that allow us to 
distinguish active and passive nodes, depending on the role they play in
the creation of edges:
\begin{itemize}
\item Some networks are directed, i.e., each edge possesses an intrinsic
  orientation from one
  node to another.  An example is an email network, in which the nodes
  are persons and directed edges represent individual email messages. 
  In such networks, the pointed-to node is the passive node and the
  pointing node is the active node. Thus, in the context of interpreting
  an edge as an attachment, it is the pointed-to node that receives the
  attachment. Therefore, we will only consider the indegree of nodes in
  these networks, i.e.,
  the number of edges pointing to a given node. 
\item Some networks are bipartite.
  A bipartite network contains two kinds of vertices,
  and all edges connect one node type with the other.  In these
  networks, we can distinguish between active and passive nodes. As an
  example, the user--song network where edges represent the ``has
  listened to'' relationship contains users (which are active) and songs
  (which are passive).  We will thus only consider the set of passive nodes
  and their degrees in this case. 
\end{itemize}

\begin{figure}
  \subfigure[Outdegrees (followees)]{
    \includegraphics[width=\wTwo]{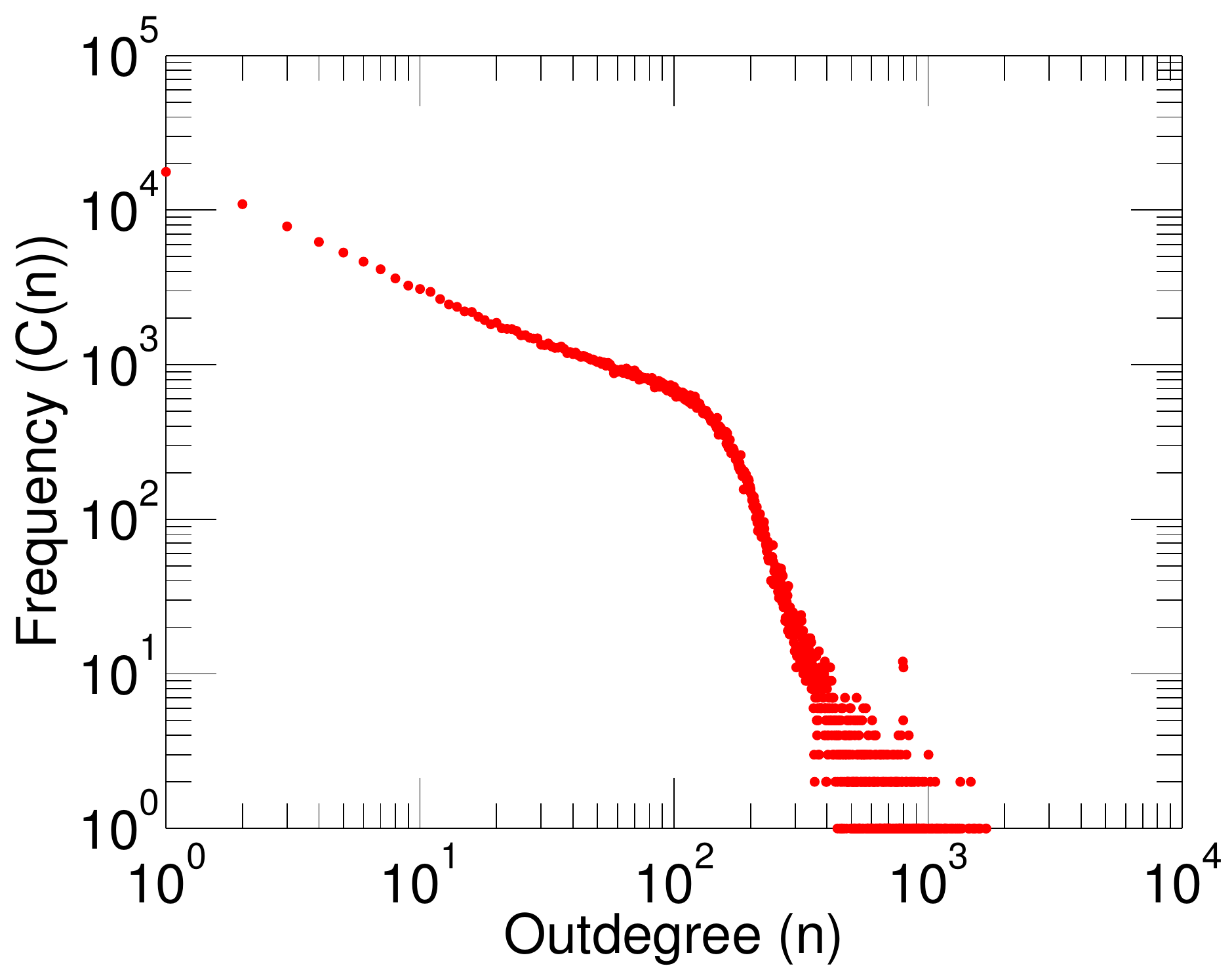}}
  \subfigure[Indegrees (followers)]{
    \includegraphics[width=\wTwo]{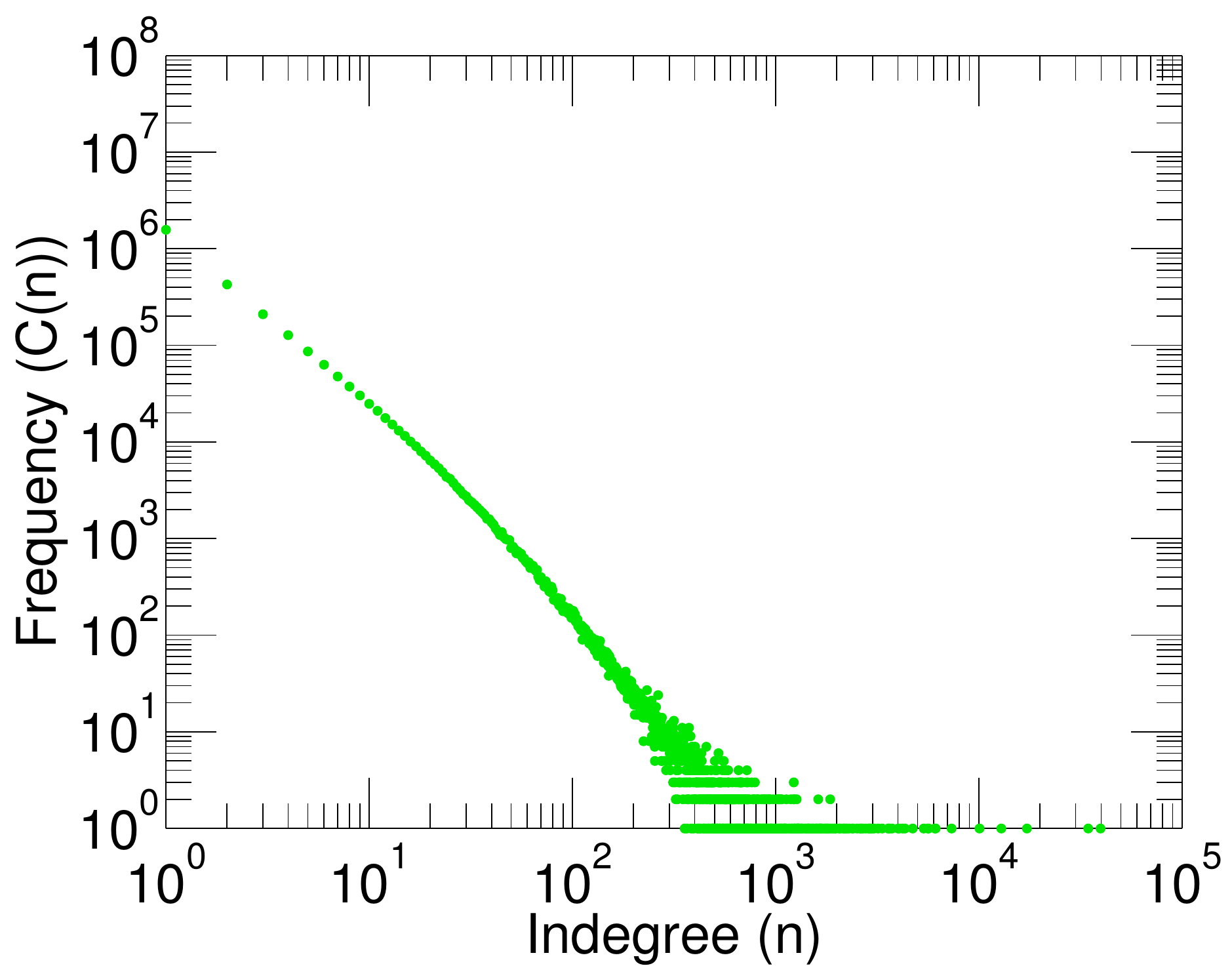}}
  \subfigure[Left degrees (users)]{
    \includegraphics[width=\wTwo]{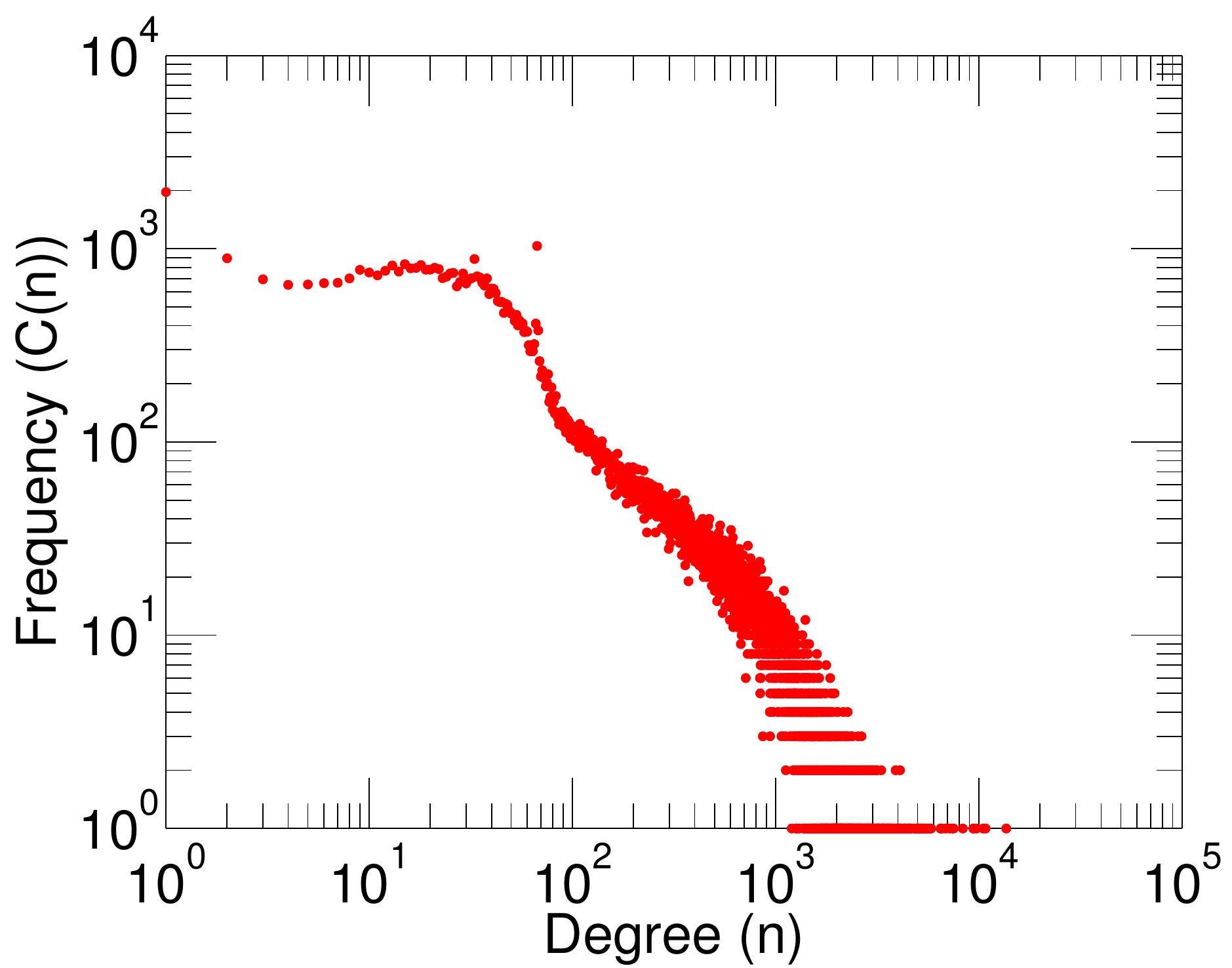}}
  \subfigure[Right degrees (movies)]{
    \includegraphics[width=\wTwo]{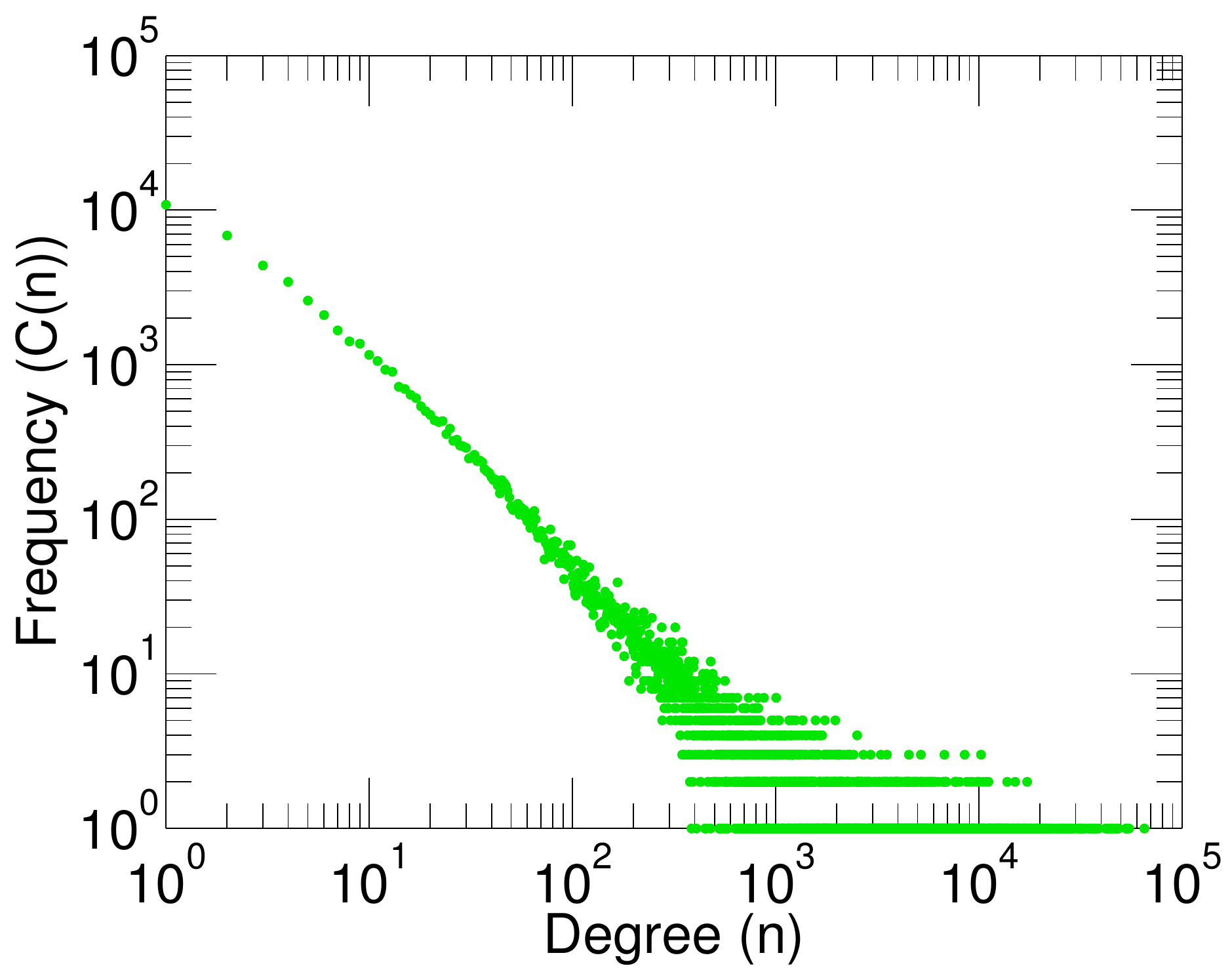}}
  \caption{
    Examples of the degree distributions of active and passive nodes in
    one directed and one bipartite network. The passive degree
    distributions are much nearer to a power law than the active degree
    distributions. 
    (a-b) The outdegree and indegree distribution of the directed social
    network of Twitter (\textsf{Wa}), 
    (c-d) The left and right degree distribution of the bipartite
    Filmtipset rating network (\textsf{Fr}). 
    (a) and (c) are active degree distribution; (b) and (d) are passive
    degree distributions. 
    \label{fig:active}
  }
\end{figure}

Note that in these two cases, the resulting degree distribution of
passive nodes tends to exhibit power-law behavior much more often than
the active nodes, as shown in Figure~\ref{fig:active} with two
examples. In the following, we describe the seven categories of networks
we study. 

\paragraph{Social networks} 
Social networks consist of persons connected by social ties
such as friendship.  The social networks we consider are based on online
social networking sites and therefore, the considered social ties are
online contacts. 
Social networks allow only a single edge between a given node pair,
i.e., multiple edges are not allowed.
Some social networks have positive and negative edges, representing
positive and negative ties such as friendship and enmity, or trust and
distrust.  In that case, we are only interested in the presence of a
tie. 
Some social networks are undirected, and others are directed.  For
directed social networks, a directed edge from person A to person B
means that person A is following or otherwise connected in an unilateral
way with another person. These directed connections can of course be
reciprocated, in which case two edges connect a given node pair in
opposite direction. 
Preferential attachment in a social networks results in the rule that
people who already have many ties are more likely to receive new
ties. In directed social networks, preferential attachment means that
people who are followed by many people (i.e., are popular) are more
likely to receive new followers. 

\paragraph{Rating networks}
A rating network is a bipartite network between persons and items they
have rated. The nodes are persons and items, and each edge connects a
person with an item, and is annotated with a rating.  
In the datasets we will consider, items can be movies, songs, products, 
jokes and even sexual escorts. 
Note that persons in rating networks are often called users, since the
datasets are used in online recommender systems. 
The ratings values in rating networks will
be ignored in our experiments. In other words, we consider only the
information whether a person has rated an item.  
Rating networks usually only allow a single rating of a given item by a
given person, although we also allow networks with parallel edges in this
category. 
We also include in this category unweighted rating-like features such as
persons liking items or marking them as favorites, since we ignore ratings
anyway; in both cases an edge represents an endorsement (positive or
negative) of an item by a person. 
We consider preferential attachment only of items.  In other words,
preferential attachment in rating networks refers to the fact that items
with many items will receive more ratings in the future. Note that this
statement is independent of the actual ratings given, but only refers to
the information of whether a rating was or will be given. 

\paragraph{Communication networks}
A communication network consists of persons which exchange information
in the form of individual messages.  Since each message is represented
by an edge, multiple edges connecting two persons are allowed.  Edges in
communication networks are always directed and can represent emails,
other types of messages in social media such as ``wall posts'' in
Facebook, or replies to another person in online forums. 
Preferential attachment in communication networks refers to the
observation that persons who have already received many messages are
more likely to receive messages in the future. 

\paragraph{Folksonomies}
Folksonomies consist of a set of tag assignments, which are
person--tag--item triples that denote that a given person has assigned a
given tag to a given item.  Items can be as diverse as websites, movies or
scientific papers. Tags are strings which are intended to describe or
classify the item. We consider two types of preferential attachment types
in folksonomies:  preferential attachment on tags and preferential
attachment on items. 
Preferential attachment on tags refers to the observation that tags
which have been used in many tag assignments are more likely to be used
in new tag assignments.  
On the other hand, preferential attachment on items refers to the
observation that items which have been tagged often are more likely to
receive tag assignments in the future. We will distinguish the two cases
by simply considering them as two different bipartite networks:  the
person--tag network and the person--item network. 

\paragraph{Wiki edit networks}
Wiki edit networks are bipartite networks between users of wikis and the
pages they edit, where each edge denotes a single edit. 
Wiki edit networks thus allow multiple edges between a user--page
combination.  
All wikis considered are Wikimedia sites such as Wikipedia, Wiktionary,
etc. 
Preferential attachment in edit networks refers to the
observation that pages which have received many edits are more likely to
receive many edits in the future.  Note that wiki edit networks are part
of the more general category of authorship networks, but that for usual
(non-wiki) works, the set of authors is fixed and thus preferential
attachment is impossible. 

\paragraph{Explicit interaction networks}
An explicit interaction network consists of people and interactions
between them. Examples are people that meet each other, or scientists
that write a paper together.  Explicit interaction networks are
unipartite and allow multiple edges, i.e., there can always be multiple
interactions between the same two persons. 
They can also be both directed or undirected. 
Preferential attachment in explicit interaction networks refers to the
observation that people who have had many interactions with other people
in the past are more likely to have interactions in the future. 

\paragraph{Implicit interaction networks}
Implicit interaction networks are networks where the interaction between
people is not encoded in direct ties between them, but in indirect ties
through things with which people interact. Thus, an implicit interaction
network is a bipartite network consisting of people and things, in which
each edge represents an interaction. Examples of implicit interaction
networks are people writing in forums, commenting on movies or listening
to a song. 
In implicit interaction networks, we always allow multiple edges between
the same person--thing pair. 
Preferential attachment in implicit interaction networks refers to the
observation that things which have been interacted with many times in the
past are more likely to receive interactions in the future. 

%% In addition to the properties described above, all networks we consider
%% have edges annotated with their creation time. 

\subsection{Definitions}
Let $G=(V,E)$ be an undirected, unweighted network allowing multiple edges, in
which $V$ is the set of nodes and $E$ is the multiset of edges. The
number of neighbors of a node $u\in V$ is called its degree and is
defined as
\begin{align*}
  d(u) &= |\{\{u,v\} \in E \mid v \in V\}|.
\end{align*}
We explicitly allow multiple edges between two nodes, and count them
separately in this definition of the degree. 
We ignore all edge weights, such as ratings in rating networks
or the positive and negative signs of signed networks. 
We also allow loops, i.e., edges from a node to itself, as these may
appear for instance in email networks when people send an email to
themselves. 

In order for preferential attachment to be observed directly (as opposed
to observing the resulting degree distribution), we need to know the
evolution of a given network. Thus, we need to know at which time each
edge was added to the network and can then consider the evolution of
the network as a function of time. 
We thus need to observe a temporal network at an intermediate time $t_1$ and at
the latest known time $t_2$.  In this paper, we will choose $t_1$ such
that at the time $t_1$, 75\% of all edges have been added to the
network.  This value is chosen such that is corresponds to the split used
in the link prediction studies between known and unknown edges
\cite{b256}.  Note 
that preferential attachment has been exploited for implementing link
prediction, under the same conditions we use here. 
For a given network $G = (V,E)$, let $G_1=(V, E_1)$ be the network
with the same vertex set as $G$, and containing all edges created before
$t_1$.  Denoting by $t(\{u,v\})$ the creation time of an edge $\{u,v\}$, we
have 
\begin{align*}
  E_1 &= \{ \{u,v\} \in E \mid t(\{u,v\}) < t_1 \}. 
\end{align*}
Let $d_1(u)$ be the degree of a node in $E_1$, i.e., 
\begin{align}
  d_1(u) &= |\{\{u,v\} \in E_1 \mid v \in V\}|. 
  \label{eq:split}
\end{align}

\subsection{Preferential Attachment Functions}
We can now give the definition of a preferential attachment function. 
A preferential attachment function is a function that maps the number of
edges at time $t_1$ to the number of news edges received
after $t_1$. 
In other words, a function $f$ such that $f(d_1(u))$
approximates $d_2(u) = d(u) - d_1(u)$ for all nodes $u \in V$. 
The values returned by a preferential attachment function $f$ will be
called attachment values. 

Different network growth models can be expressed in terms of the
preferential attachment function $f$ they are based on. 
We will consider all functions only up to a constant factor, since we
are only interested in the relative attachment values of different
vertices. 

\begin{itemize}
  \item $f(d) = 1$.  In this model, the attachment is independent of the
    degree. This growth models leads to networks in which all edges are
    equally likely independently from each other, i.e., the Erdős--Rényi
    model \cite{b569}.  These networks are usually simply called random graphs. 
  \item $f(d) = d$. In this model, preferential attachment is
    linear. This corresponds to the Barabási--Albert model of
    scale-free networks \cite{b439}. 
  \item $f(d) = d^\beta$. In this model, the attachment is an arbitrary power of
    the degree \cite{b764,b756}. 
  \item $f(d) = (1+d)^\beta$. This model modifies the previous one in
    that it gives a nonzero attachment value even to nodes of zero
    degree \cite{b767}. 
\end{itemize}

In this paper, we will use the slightly more general form
\begin{align*}
  f(d) &= e^\alpha (1+d)^\beta - \lambda \\
  &= e^{\alpha + \beta \ln (1+d)} - \lambda,
\end{align*}
in which $\lambda$ is a regularization parameter, whose purpose will
become clear in the following.
The exponent $\beta$ will be called the preferential attachment
exponent. The parameter $\alpha$ is a multiplicative term that we can
ignore since values of $f$ are to be interpreted only up to a constant
factor. 

\subsection{Generalization of Previous Models}
Individual graph growth
models can be recovered by setting the parameter $\beta$ in the
preferential attachment function $f(d) = d^\beta$ to specific
values.  

(a) \textbf{Constant case $\beta=0$}.  
This case is equivalent to a constant function $f(d)$, 
and thus this graph growth model results in networks in which each edge
is equally likely and independent from other edges.  This is the
Erdős--Rényi model of random graphs \cite{b569}. 

(b) \textbf{Sublinear case $0 < \beta < 1$}. 
In this case, the preferential attachment function is sublinear.  This
model gives rise to a stretched exponential degree distribution
\cite{b764}, whose exact 
expression is complex and given in \cite[Eq. 94]{b773}. 

(c) \textbf{Linear case $\beta=1$}.
This is the scale-free network model of Barabási and Albert
\cite{b439}, in which attachment is proportional to the degree. This gives a
power law degree distribution.  

(d) \textbf{Superlinear case $\beta > 1$}.
In this case, a single node will acquire 100\% of all edges
asymptotically \cite{b765}. Networks with this behavior will however display
power law degree distributions in the pre-asymptotic regime
\cite{b769}. 

\subsection{Curve Fitting}
We now describe our method for estimating the 
value of the parameter $\beta$.  Since the values of the degree $d$ span
several orders of magnitude, a simple least-squares curve fitting would
give highly skewed results, as it would drastically over-weight high
degrees. Therefore, we perform a least square fitting on the logarithmic
degrees.
The following minimization problem gives an estimate for the exponent
$\beta$.
\begin{align}
  \min_{\alpha,\beta} \sum_{u\in V} \left( \alpha + \beta \ln[1 + d_1(u)]
  - \ln[\lambda + d_2(u)]  \right)^2
  \label{eq:min}
\end{align}
The resulting value of $\beta$ is the estimated preferential attachment
exponent. 
Note that due to the shift term of one and the regularization parameter
$\lambda$, our model can both accommodate nodes with degree zero at time
$t_1$, as well as nodes that do not receive any new link between $t_1$
and $t_2$. 

To measure the error of the fit, we define the root-mean-square
logarithmic error $\epsilon$ in the following way:
  \small
\begin{align*}
    \epsilon &= \exp\left\{ \sqrt{ \frac 1 {|V|} \sum_{u \in V}
      \left(\alpha + \beta \ln[1 
        + d_1(u)] - \ln[\lambda + d_2(u)]\right)^2  }  \right\}
\end{align*}
\normalsize
This gives the average factor by which the actual new number of edges
differs from the predicted value, computed logarithmically. 
The value of $\epsilon$ is larger or equal to one by construction. 

\section{Experiments}
We compute an estimation of the preferential attachment exponent $\beta$
for forty-seven network datasets. 
All networks are taken from the Koblenz Network Collection (KONECT,
\href{http://konect.uni-koblenz.de/}{konect.uni-koblenz.de}). 
The full description of the networks can be read on the KONECT
website\footnote{\href{http://konect.uni-koblenz.de/networks}{konect.uni-koblenz.de/networks}}.    
The networks we use in our experiments fulfill the following criteria:
\begin{itemize}
\item Creation times are known for all edges. 
\item We exclude very incomplete datasets, in which the degree
  distributions are skewed by the sampling method used to generate the
  data.  
  %% The concerned networks are 
  %% from Twitter (tags) and Slashdot (threads). 
\item We exclude networks that are too small, i.e., have less than
  10,000 edges. 
  %% (elec, sociopatterns-infectious/hypertext are not included)
\end{itemize}

Table \ref{tab:networks} shows the list of datasets used in our
experiments. 

\begin{table}
  \caption{
    The list of network datasets used in this paper.
    Flags:  U = Unipartite, D = Directed, B = Bipartite, M = Multiple
    edges.  $|V|$ refers to the number of nodes in the network; $|E|$
    refers to the number of edges in the network. 
    In all networks, edges are annotated with edge creation times. 
  }
  \label{tab:networks}
  \centering
  \input{networks}
\end{table}

\subsection{Methodology}
For each network, we split the set of edges into the set of old edges
$E_1$ and the set of new edges $E \setminus E_1$ 
as described in Equation~\eqref{eq:split}.  Then, we compute the old
degree $d_1(u)$ for all nodes $u\in V$ and the number of new edges
$d_2(u) = d(u) - d_1(u)$. We then solve the least-squares minimization problem
of Equation~\eqref{eq:min}, giving an estimate of the preferential
attachment exponent $\beta$. 

The regularization parameter $\lambda$ is set to $0.1$ in our
experiments. 

\subsection{Experimental Results}
\begin{figure*}[t]
  \subfigure[Flickr]{
    \includegraphics[width=\wFullThree]{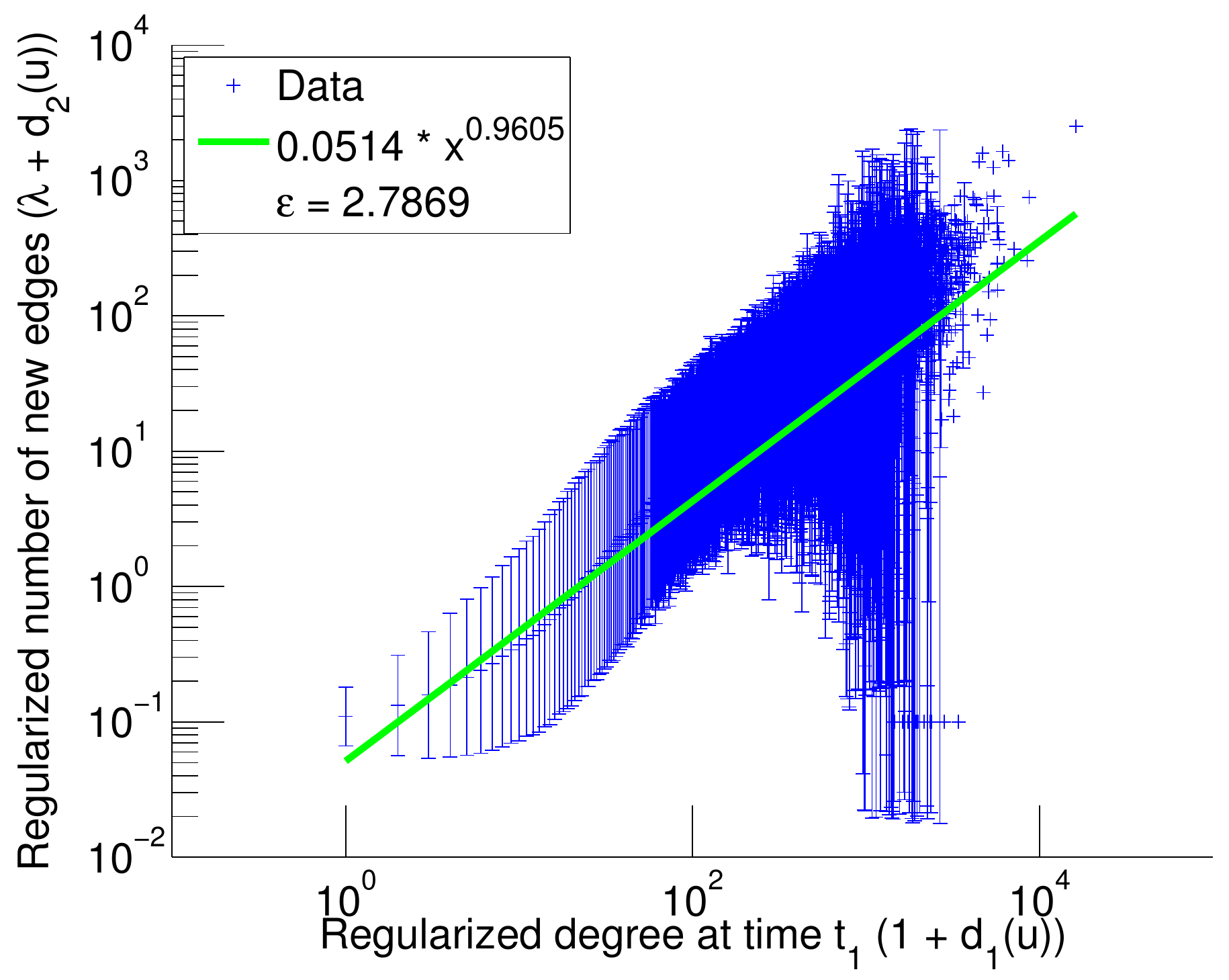}}
  \subfigure[Yahoo Songs]{
    \includegraphics[width=\wFullThree]{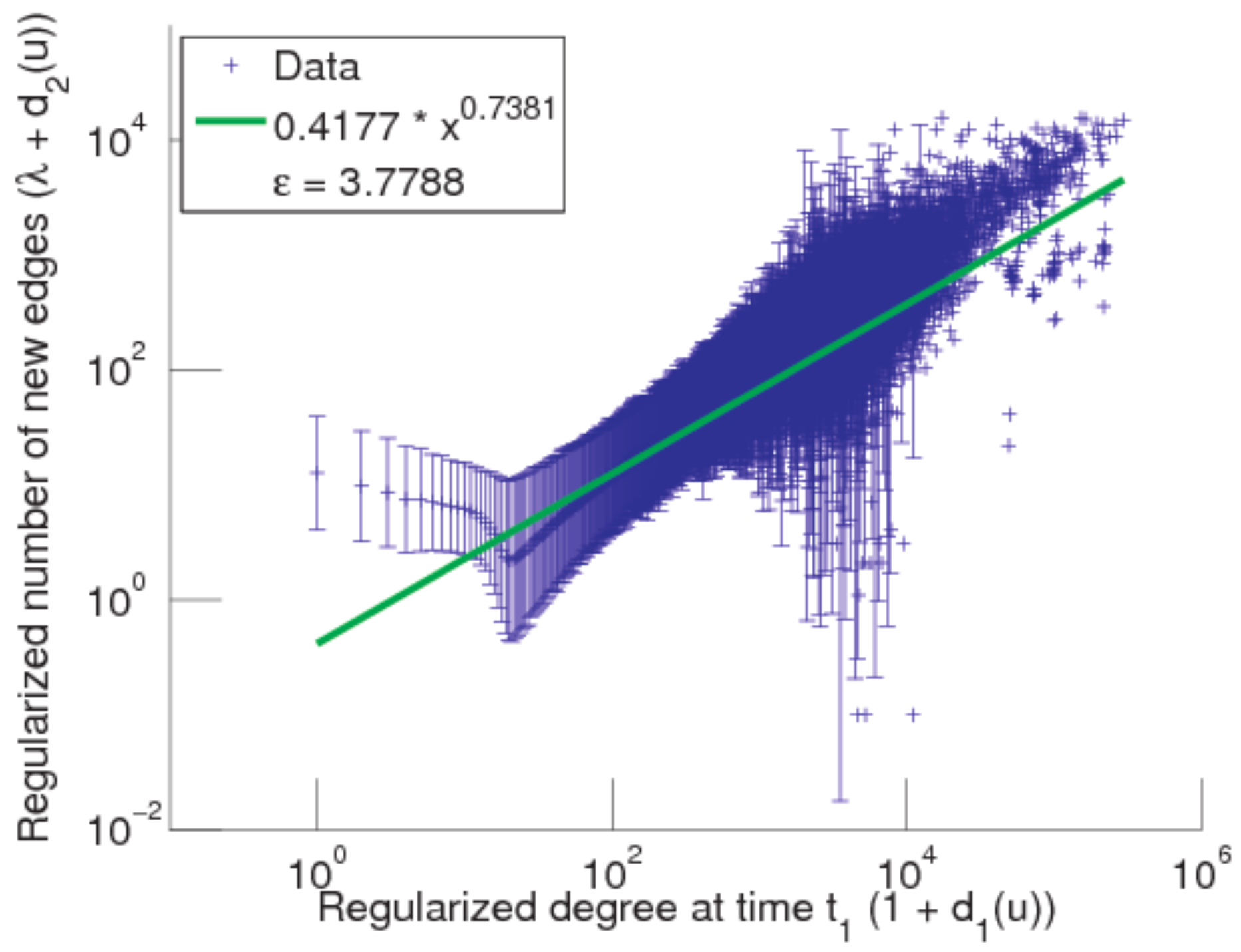}}
  \subfigure[Enron]{
    \includegraphics[width=\wFullThree]{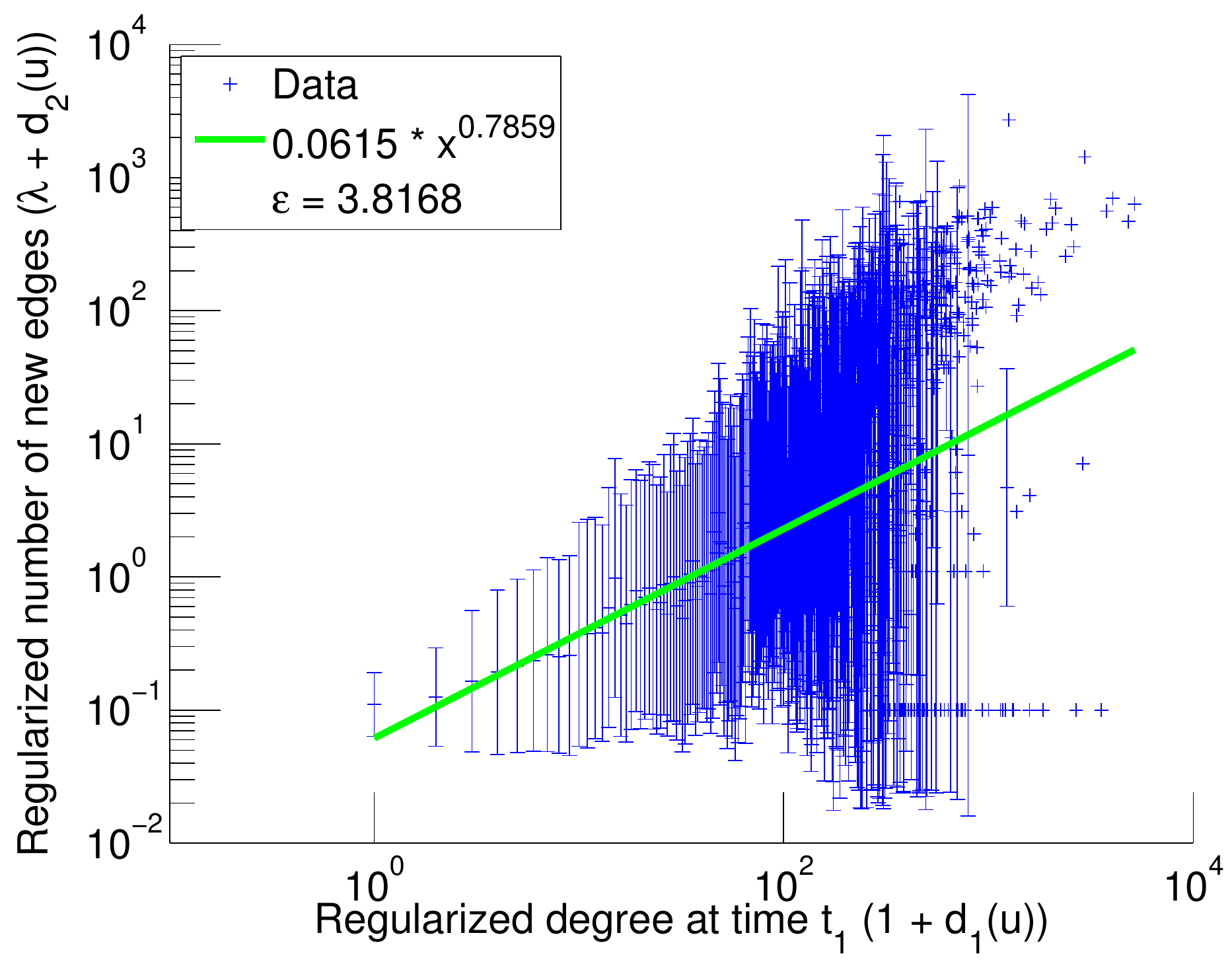}}
  \subfigure[CiteULike user--publication]{
    \includegraphics[width=\wFullThree]{pa_kv_enron}}
  \subfigure[English Wikipedia]{
    \includegraphics[width=\wFullThree]{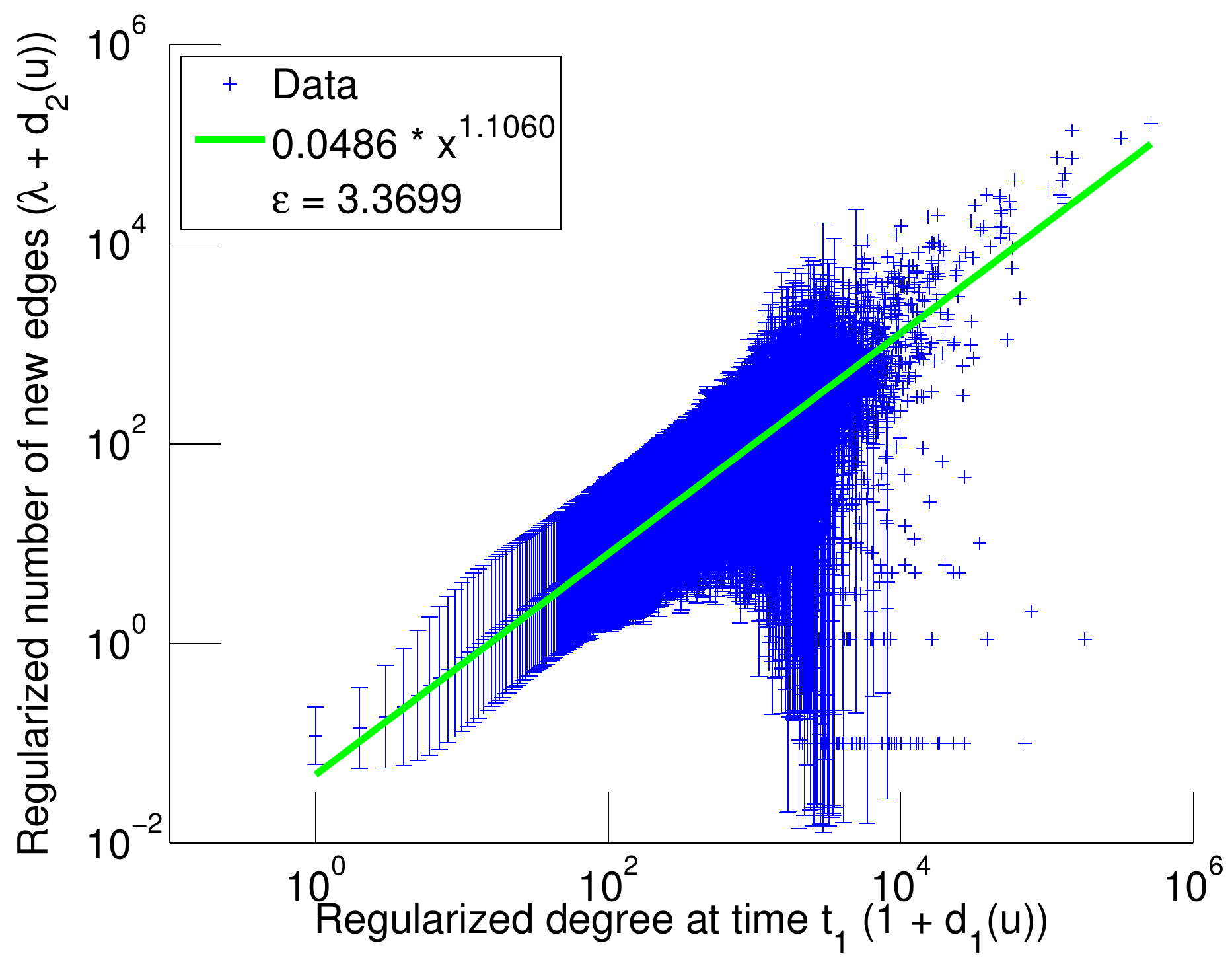}}
  %% \subfigure[DBLP]{
  %%   \includegraphics[width=\wFullThree]{plot/pa.ka.dblp_coauthor}}
  \subfigure[Last.fm song]{
    \includegraphics[width=\wFullThree]{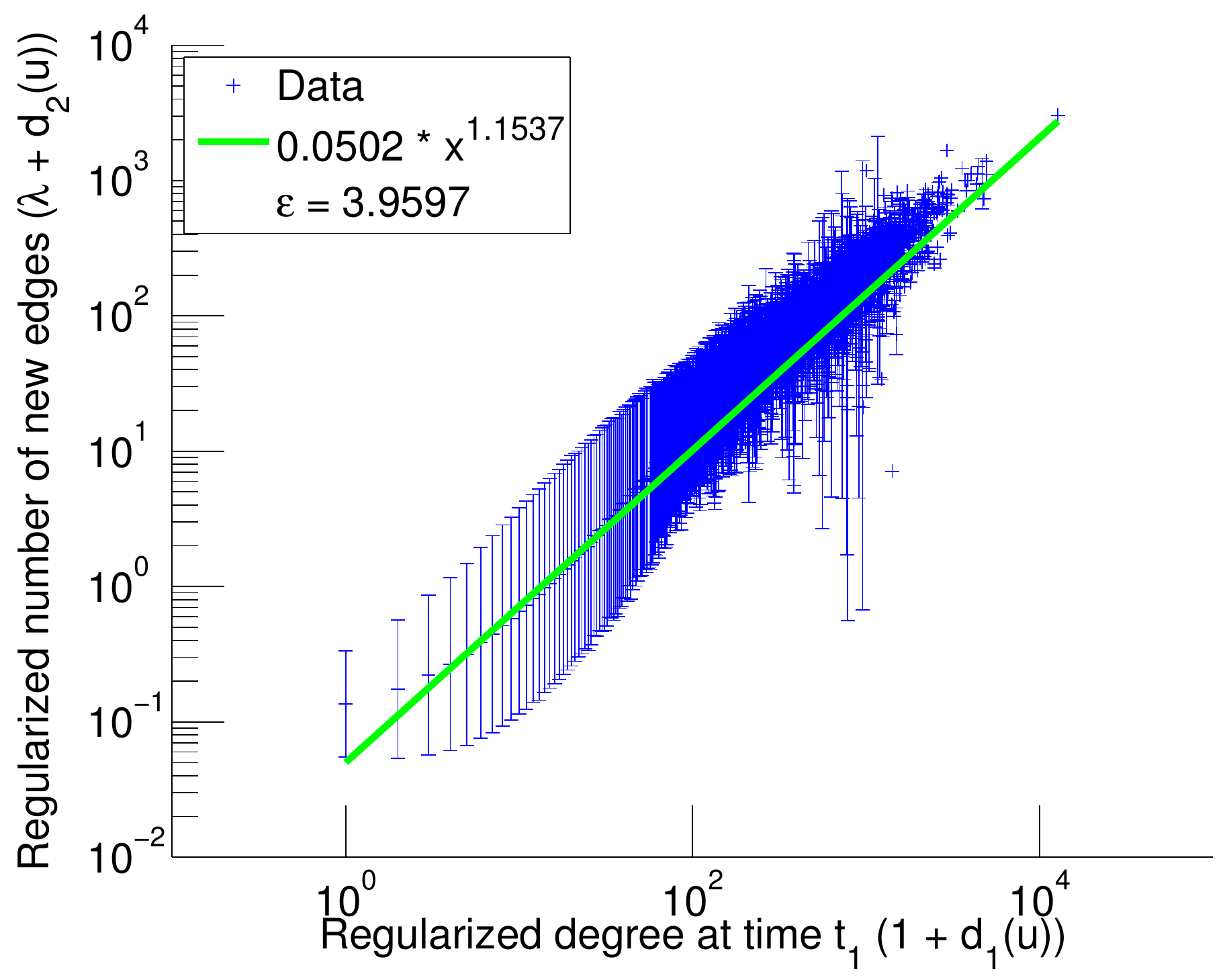}}
  \caption{
    The mean number of new edges $d_2(u) = d(u)-d_1(u)$ as a function of
    $d_1(u)$ in the largest networks of six network categories. 
    (a)~Epinions trust (\textsf{EP}), a social network 
    (b)~Yahoo Songs (\textsf{YS}), a rating network
    (c)~Enron (\textsf{EN}), a communication network
    (d)~CiteULike user--publication (\textsf{Cui}), a folksonomy,
    (e)~English Wikipedia (\textsf{en}), a wiki edit network,
    %% (e)~DBLP (\textsf{Pc}), an explicit interaction network, 
    (f)~Last.fm song (\textsf{Ls}), a implicit interaction network. 
    The bars indicate the logarithmic standard deviation
    measured over all nodes with the same value of $d_1(u)$. 
    The line represents the fitting curve $f(d) = e^\alpha
    (1+d_1)^\beta-\lambda$. 
    The standard deviation and mean on the plot is shown for
    illustration; it is not used in the fitting procedure.  The actual
    curve fitting is performed by solving the 
    optimization problem in Equation \ref{eq:min}. 
    For small values of $d_1(u)$, the standard deviation is small due
    to the high number of nodes having low degree. For larger values of
    $d_1(u)$, the standard deviation becomes higher due to the
    reduced number of nodes with high degree. For very large values of
    $d_1(u)$, only one node has a given degree, and the
    standard deviation is undefined.  
    \label{fig:pa.k}
  }
\end{figure*}

To illustrate the curve fitting procedure, we show the mean standard
deviation of the logarithmic degree $d_2 = d(u) - d_1(u)$ as a function of
$d_1(u)$. Figure \ref{fig:pa.k} shows this plot, along with the
fitted curve, for the largest networks of six different network
categories.

\begin{figure}
  \centering
  \includegraphics[width=\wOne]{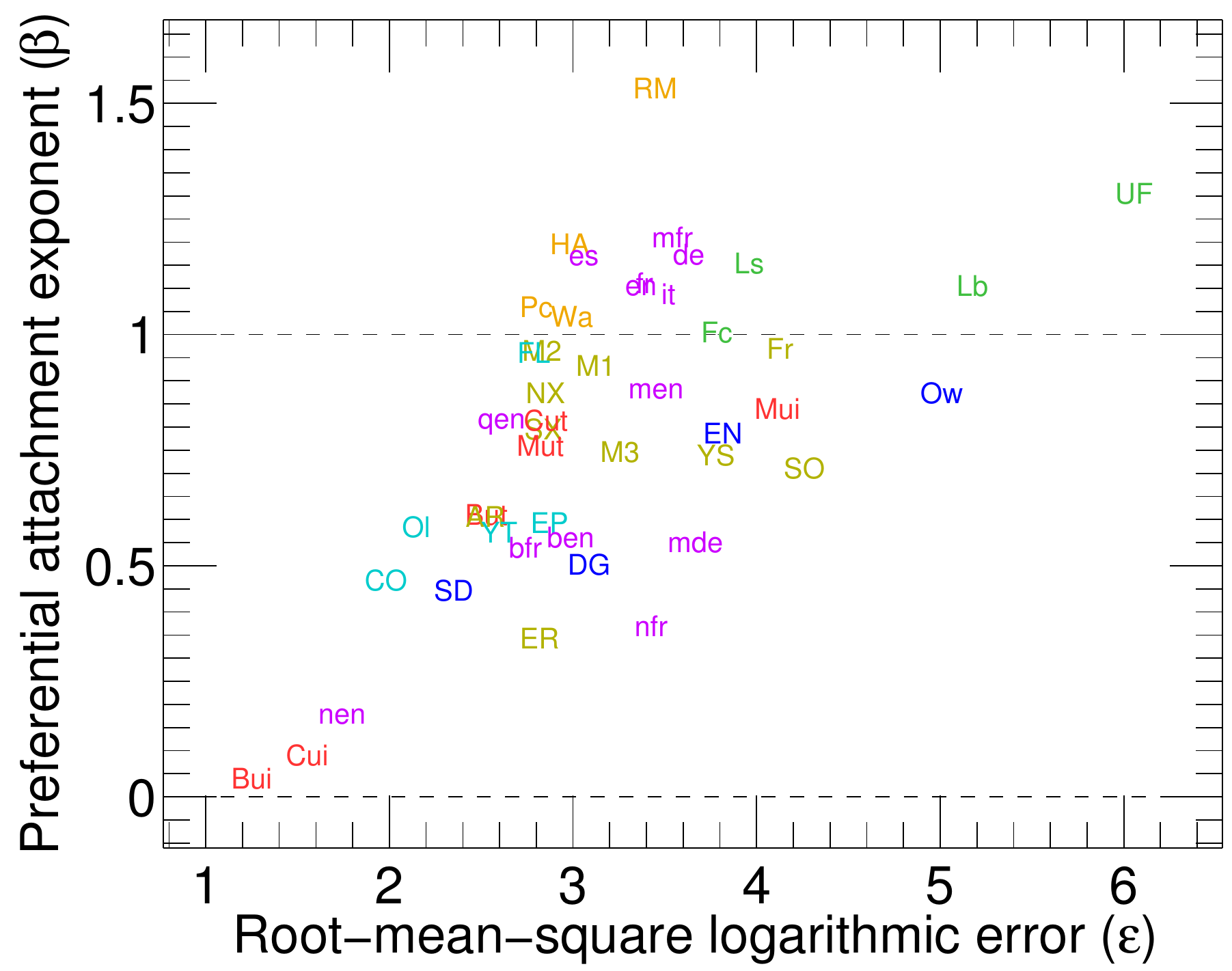}
  \includegraphics[width=0.35\textwidth]{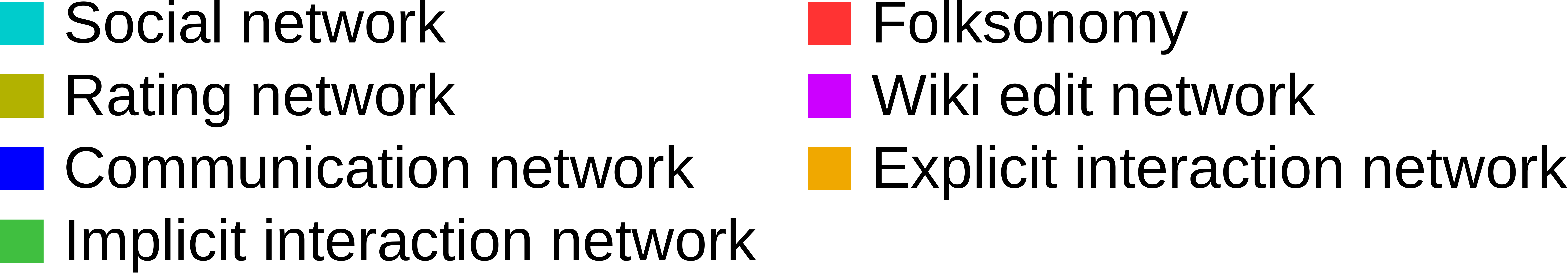}
  \caption{
    The preferential attachment exponent $\beta$ and the root-mean-square logarithmic error $\epsilon$. 
    Each two or three letter code represents one network dataset. The
    codes are given in Table~\ref{tab:networks}.
    The color of the codes represent the network category. 
    \label{fig:scatter.mse}
  }
\end{figure}

The root-mean-square logarithmic error $\epsilon$ is shown together with
$\beta$ in 
Figure~\ref{fig:scatter.mse}. 

\begin{figure}
  \centering
  \includegraphics[width=\wOne]{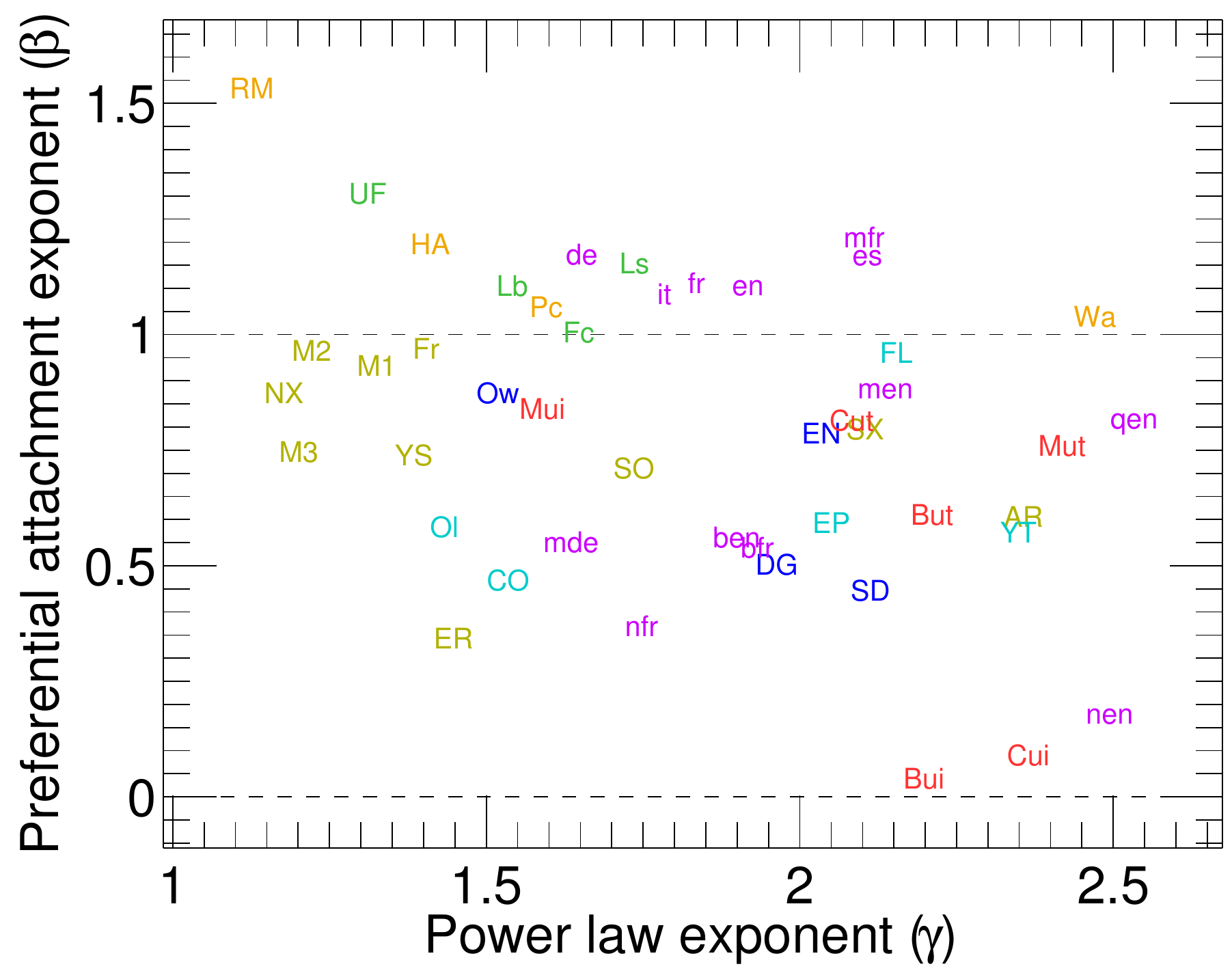}
  \includegraphics[width=0.35\textwidth]{legend-category}
  \caption[*]{
    The preferential attachment exponent $\beta$ plotted against the
    power law exponent $\gamma$.  
    Each two or three letter code represents one network dataset. 
    The
    codes are given in Table~\ref{tab:networks}.
    The color of the codes represent the network category. 
    The estimates for the power law exponent $\gamma$ are computed using the
    robust method given in \cite[Eq. 5]{b408}. 
    \label{fig:scatter.power}
  }
\end{figure}

The measured preferential attachment exponents $\beta$ for all networks
are shown
in Figure~\ref{fig:scatter.power}. 
The estimates for the power law exponent $\gamma$ are computed using the
robust method given in \cite[Eq. 5]{b408}. 
We note that the estimated degree
distribution exponents lie in the approximate range $[1,2.5]$, and are
thus smaller than the usually cited range $[2,3]$ would suggest. 

\begin{figure}
  \centering
  \includegraphics[width=\wOne]{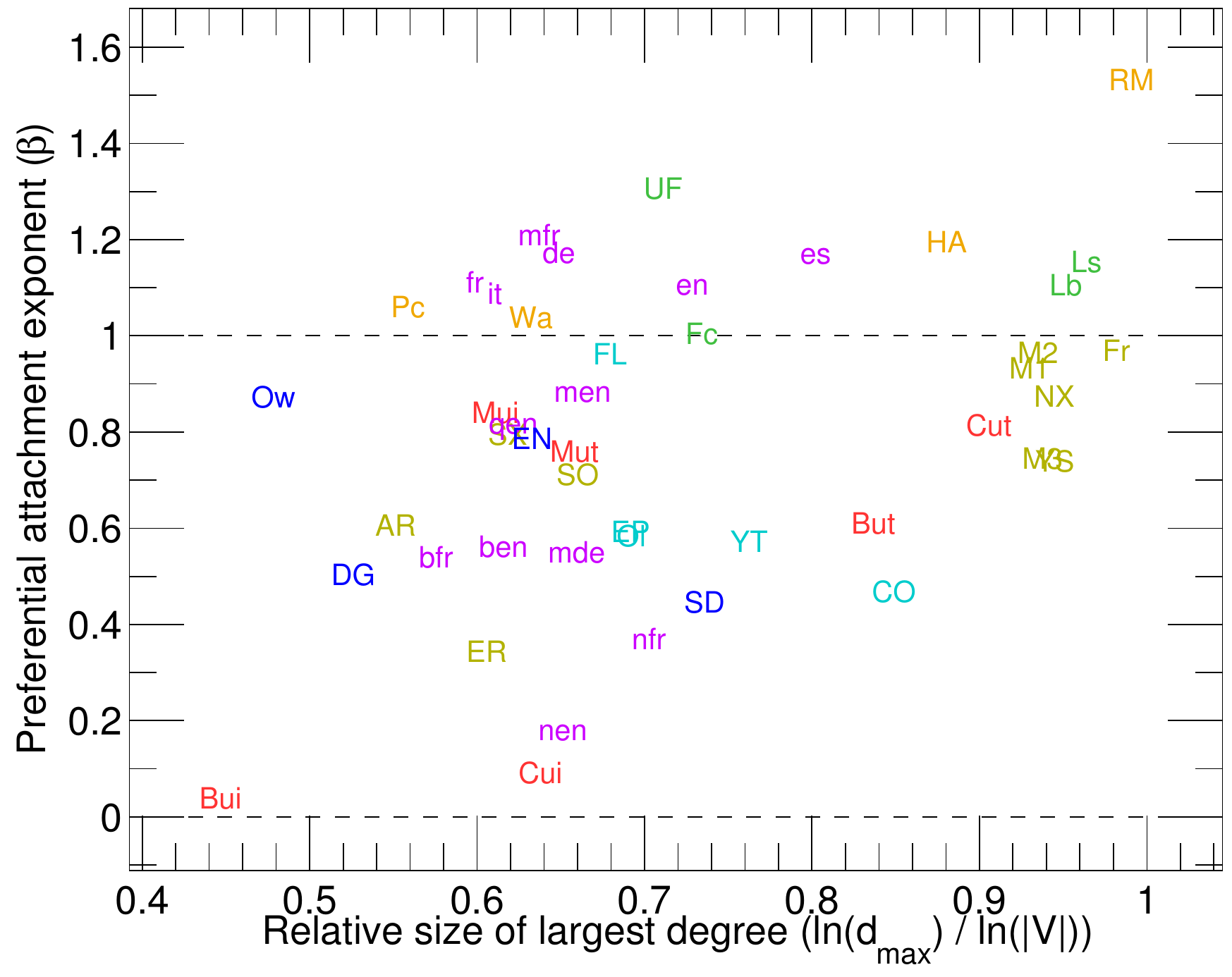}
  \includegraphics[width=0.35\textwidth]{legend-category}
  \caption{
    The relative size of the largest degree in the network plotted against the
    estimated preferential attachment exponent $\beta$ for all
    networks. 
    Each two or three letter code represents one network dataset. 
    The
    codes are given in Table~\ref{tab:networks}.
    The color of the codes represent the network category. 
    \label{fig:scatter.patest}
  }
\end{figure}

In the case of superlinear attachment, the degree distribution is
predicted to converge over time to a state in which a single node
dominates all other nodes, i.e., in which a single node has 100\% of
all inlinks asymptotically.  
Let $d_{\max}$ be the degree of the node with most links in the networks. 
Then, to test whether such nodes are present in
the studied networks, Figure~\ref{fig:scatter.patest} shows the ratio
$\ln(d_{\max})/\ln(|V|)$ plotted against the preferential attachment
$\beta$.  
The plots exhibit a moderate agreement of the super- versus the
sublinear cases 
for networks in which $1.3 < \beta < 1.5$ (such as \textsf{RM},
\textsf{TH}, \textsf{Ls}, \textsf{Lb}, \textsf{PH}, \textsf{HA}) and 
for $0.4 < \beta < 0.7$ (like \textsf{Cui}, \textsf{Bui}, \textsf{nen},
\textsf{ER}, \textsf{DG},  \textsf{AR}, \textsf{nfr}). If $\beta$
is close one (i.e., in the case of weak sublinear or weak superlinear
attachment), the agreement breaks down.

\section{Discussion}
We have empirically investigated preferential attachment in
forty-seven online networks and found that these networks follow a
nonlinear  
preferential attachment model, i.e., their preferential attachment
exponent is either larger than one (superlinear) or lower than one
(sublinear). As such, we 
challenge the often implicit assumption in Web Science that preferential
attachment assumes a linear relationship
(cf.\ \cite{b439}). Furthermore, we show that certain clearly distinct
categories of  
online networks feature a superlinear preferential attachment exponent,
whereas other categories feature a sublinear one. Our findings point out
that previous studies of preferential attachment in online networks
might have oversimplified the underlying mechanisms by assuming
linearity when, in fact, most online networks follow a nonlinear
pattern. 

In particular, we find that the majority (70\%) of the studied online networks
fall into the sublinear category, having $\beta <
1$. The networks in the sublinear category were previously classified as
rating, communication, folksonomy and social networks (see also Table
1). Also, a subset of the authorship networks falls into this category,
specifically all wiki edit networks except those from Wikipedia (with
the exception of 
the French Wiktionary). The other 30\% fall into the superlinear
category where $\beta > 1$. These networks were classified as explicit
and implicit interaction
networks. Also, the second subset of the authorship networks falls into
this category, specifically all Wikipedias and the French Wiktionary.  

Our findings show that online networks do not follow a linear
preferential attachment model. Actually, not one of the studied networks
featured linearity where $\beta = 1$ exactly. This is unexpected, as
most literature 
implies such linearity. In addition, we show that online networks are
also not consistent in their (non-)linearity: most networks follow a
sublinear preferential attachment model, whereas others follow a
superlinear model. However, we do find patterns that suggest an
underlying internal consistency, because most of the previous
classifications fit in their entirety into one category (except the
wiki edit networks). For example, all networks in the `rating
networks' classification are part of the sublinear category, whereas all
interaction networks fall into the superlinear category.  

Furthermore, we observe that similar to the distribution of the
preferential attachment exponent, the power law exponent too is far
from consistent as suggested by previous literature \cite{b59}. We find
that while the values differ quite extremely,
the distribution does not seem to follow a distinct pattern, nor is
there a clear correlation with the preferential attachment exponent (see
Figure \ref{fig:scatter.power}).  

An explanation of these findings might be that
some networks, in particular the ones falling into the superlinear
category, follow other or at least additional governance mechanisms than
other networks \cite{b782}. For example, authors on
Wikipedia (which is part of the 
superlinear category) might perceive the existence of a strong internal
normative system that prescribes their online behavior
\cite{b783}. Indeed, previous research has often observed the 
existence of social norms and normative systems in online networks
 (e.g., 
\cite{b784,b785,b786,b787}). This phenomenon is often explained by the absence of more 
formal and explicit governance mechanisms (e.g., \cite{b788,b789}) that
are typically observed in  
other types of networks.  

\section{Conclusion}

The findings presented in this paper show that interaction in online
networks might be more complex than previously thought. In particular,
we show that these networks follow a nonlinear preferential attachment
model, contrary to what is suggested in the literature. Similarly, most
of the networks that we studied have a power law exponent that is not
even close to being consistently in the range $\gamma \in [2,3]$
\cite{b59}. This leads to the 
conclusion that research into online networks might need to take into
account other factors, and most importantly employ different models that
allow for the nonlinearity of the preferential attachment model. Also,
the previous assumption of a more or less generalized range of $\gamma
\in [2,3]$
for the power law exponent seems to be challenged, as we observed
variation in that value across the networks.  

Our work suggests a number of future research directions. First, as a
direct consequence of our empirical findings, we suggest that future
work should develop new models to allow for nonlinearity of the
preferential attachment exponent, as well as diversity of the power law
exponent. Our findings undermine many of the previously developed
models, as such we expect fruitful research in that direction. Second,
and related to our first point, future research should search for
explanations for our findings. In the previous section, we tentatively
highlight some possible explanations; however, empirical studies need to
establish their value. In particular, research should connect more
mathematical approaches to study online networks (such as presented in
this paper) and sociological attempts to explain the observed
phenomena. For example, if the type of governance in a given network
indeed influences its preferential attachment and power law exponent,
how exactly does that mechanism work? Related to this question it is
important to investigate the emergence of networks, something that we
did neglect in the current research. If we assume sociological
mechanisms to play a role in the explanation of diversity and
nonlinearity of the two exponents, then it follows that the antecedents
of these mechanisms need to be investigated. For example, how does a
certain type of governance in an online network come into being? What
are the driving forces behind this emergence, and how can these
mechanisms best be studied? We hope that our paper contributes to fuel
research into that direction. 

Our study is subject to a number of limitations that present
opportunities for future research. First, we do find that the
preferential attachment exponent is nonlinear, similarly the power law
exponent is distributed more diversely than expected. However, we do not
investigate the relationship between these two observations, and suggest
that future work further delves into this issue. Second, we investigated
forty-seven datasets. Future research might broaden the scope of our study to
include more and more diverse online networks. For example, in the
current study we did not investigate networks such as hyperlink networks and
affiliation networks. It should be fruitful to test the observations
that we make in this paper on a larger scale, and as such generalize our
findings to a broader level. 

\section{Acknowledgments}
We thank Hans Akkermans, Rena Bakhshi and Julie Birkholz for helpful
discussions and 
hints on work related to this paper.  The research leading to the
results presented in this paper has received funding from the
European Community's Seventh Frame Programme under grant agreement
n\textsuperscript{o}~257859, \href{http://robust-project.eu/}{ROBUST}.

\balance

% If you want to use smaller typesetting for the reference list,
% uncomment the following line:
 \small
\bibliographystyle{acm-sigchi}
\bibliography{ref,preferential-attachment,konect,ukob}
\end{document}

%% file: networks.tex
\definecolor{colorSocial}{rgb}{0.000000,0.800000,0.800000}
\definecolor{colorRatings}{rgb}{0.700000,0.700000,0.000000}
\definecolor{colorCommunication}{rgb}{0.000000,0.000000,1.000000}
\definecolor{colorFolksonomy}{rgb}{1.000000,0.200000,0.200000}
\definecolor{colorAuthorship}{rgb}{0.800000,0.000000,1.000000}
\definecolor{colorContact}{rgb}{0.940000,0.660000,0.000000}
\definecolor{colorInteraction}{rgb}{0.250000,0.750000,0.250000}
 \scalebox{0.8}{ \begin{tabular}{@{} r @{\;} l @{\;} l l @{} r @{\;} r @{}}\toprule & \textcolor{colorSocial}{$\blacksquare$}  & \textbf{Social networks} & \textbf{Flags} & \textbf{$|V|$} & \textbf{$|E|$} \\ \midrule
\cite{b367} & \textsf{\href{http://konect.uni-koblenz.de/networks/epinions}{EP}} & \href{http://konect.uni-koblenz.de/networks/epinions}{Epinions trust} & D & 131,828 & 841,372 \\
\cite{b480} & \textsf{\href{http://konect.uni-koblenz.de/networks/facebook-wosn-links}{Ol}} & \href{http://konect.uni-koblenz.de/networks/facebook-wosn-links}{Facebook friendships} & U & 63,731 & 1,545,686 \\
\cite{konect:brandes09} & \textsf{\href{http://konect.uni-koblenz.de/networks/wikiconflict}{CO}} & \href{http://konect.uni-koblenz.de/networks/wikiconflict}{Wikipedia conflict} & U & 118,100 & 2,985,790 \\
\cite{b518} & \textsf{\href{http://konect.uni-koblenz.de/networks/youtube-u-growth}{YT}} & \href{http://konect.uni-koblenz.de/networks/youtube-u-growth}{YouTube} & D & 3,223,643 & 18,524,095 \\
\cite{b494} & \textsf{\href{http://konect.uni-koblenz.de/networks/flickr-growth}{FL}} & \href{http://konect.uni-koblenz.de/networks/flickr-growth}{Flickr} & D & 2,302,925 & 33,140,018 \\
\bottomrule\toprule & \textcolor{colorRatings}{$\blacksquare$}  & \textbf{Rating networks} & \textbf{Flags} & \textbf{$|V|$} & \textbf{$|E|$} \\ \midrule
\cite{konect:Rocha2010} & \textsf{\href{http://konect.uni-koblenz.de/networks/escorts}{SX}} & \href{http://konect.uni-koblenz.de/networks/escorts}{Sexual escorts} & B & 16,730 & 50,632 \\
\cite{www.grouplens.org/node/73} & \textsf{\href{http://konect.uni-koblenz.de/networks/movielens-100k_rating}{M1}} & \href{http://konect.uni-koblenz.de/networks/movielens-100k_rating}{MovieLens 100k} & B & 2,625 & 100,000 \\
\cite{www.grouplens.org/node/73} & \textsf{\href{http://konect.uni-koblenz.de/networks/movielens-1m}{M2}} & \href{http://konect.uni-koblenz.de/networks/movielens-1m}{MovieLens 1M} & B & 9,746 & 1,000,209 \\
\cite{konect:stackexchange} & \textsf{\href{http://konect.uni-koblenz.de/networks/stackexchange-stackoverflow}{SO}} & \href{http://konect.uni-koblenz.de/networks/stackexchange-stackoverflow}{Stack Overflow} & B & 641,876 & 1,302,439 \\
\cite{konect:lim2010} & \textsf{\href{http://konect.uni-koblenz.de/networks/amazon-ratings}{AR}} & \href{http://konect.uni-koblenz.de/networks/amazon-ratings}{Amazon ratings} & B & 3,376,972 & 5,838,041 \\
\cite{www.grouplens.org/node/73} & \textsf{\href{http://konect.uni-koblenz.de/networks/movielens-10m_rating}{M3}} & \href{http://konect.uni-koblenz.de/networks/movielens-10m_rating}{MovieLens 10M} & B & 80,555 & 10,000,054 \\
\cite{b367} & \textsf{\href{http://konect.uni-koblenz.de/networks/epinions-rating}{ER}} & \href{http://konect.uni-koblenz.de/networks/epinions-rating}{Epinions product ratings} & B & 876,252 & 13,668,320 \\
\cite{said:social-similarity} & \textsf{\href{http://konect.uni-koblenz.de/networks/filmtipset_rating}{Fr}} & \href{http://konect.uni-koblenz.de/networks/filmtipset_rating}{Filmtipset} & B & 144,671 & 19,554,219 \\
\cite{b520} & \textsf{\href{http://konect.uni-koblenz.de/networks/netflix}{NX}} & \href{http://konect.uni-koblenz.de/networks/netflix}{Netflix} & B & 497,959 & 100,480,507 \\
\cite{yahoo-song} & \textsf{\href{http://konect.uni-koblenz.de/networks/yahoo-song}{YS}} & \href{http://konect.uni-koblenz.de/networks/yahoo-song}{Yahoo Songs} & B & 1,625,951 & 256,804,235 \\
\bottomrule\toprule & \textcolor{colorCommunication}{$\blacksquare$}  & \textbf{Communication networks} & \textbf{Flags} & \textbf{$|V|$} & \textbf{$|E|$} \\ \midrule
\cite{konect:opsahl09} & \textsf{\href{http://konect.uni-koblenz.de/networks/opsahl-ucsocial}{UC}} & \href{http://konect.uni-koblenz.de/networks/opsahl-ucsocial}{UC Irvine messages} & D M & 1,899 & 59,835 \\
\cite{b565} & \textsf{\href{http://konect.uni-koblenz.de/networks/munmun_digg_reply}{DG}} & \href{http://konect.uni-koblenz.de/networks/munmun_digg_reply}{Digg} & D M & 30,398 & 87,627 \\
\cite{konect:slashdot-threads} & \textsf{\href{http://konect.uni-koblenz.de/networks/slashdot-threads}{SD}} & \href{http://konect.uni-koblenz.de/networks/slashdot-threads}{Slashdot threads} & D M & 51,083 & 140,778 \\
\cite{b480} & \textsf{\href{http://konect.uni-koblenz.de/networks/facebook-wosn-wall}{Ow}} & \href{http://konect.uni-koblenz.de/networks/facebook-wosn-wall}{Facebook wall posts} & D M & 63,891 & 876,993 \\
\cite{b345} & \textsf{\href{http://konect.uni-koblenz.de/networks/enron}{EN}} & \href{http://konect.uni-koblenz.de/networks/enron}{Enron} & D M & 87,273 & 1,148,072 \\
\bottomrule\toprule & \textcolor{colorFolksonomy}{$\blacksquare$}  & \textbf{Folksonomies} & \textbf{Flags} & \textbf{$|V|$} & \textbf{$|E|$} \\ \midrule
\cite{www.grouplens.org/node/73} & \textsf{\href{http://konect.uni-koblenz.de/networks/movielens-10m_ui}{Mui}} & \href{http://konect.uni-koblenz.de/networks/movielens-10m_ui}{MovieLens user–movie} & B M & 11,610 & 95,580 \\
\cite{www.grouplens.org/node/73} & \textsf{\href{http://konect.uni-koblenz.de/networks/movielens-10m_ut}{Mut}} & \href{http://konect.uni-koblenz.de/networks/movielens-10m_ut}{MovieLens user–tag} & B M & 20,537 & 95,580 \\
\cite{konect:bibsonomy} & \textsf{\href{http://konect.uni-koblenz.de/networks/bibsonomy-2ut}{But}} & \href{http://konect.uni-koblenz.de/networks/bibsonomy-2ut}{BibSonomy user–tag} & B M & 210,467 & 2,555,080 \\
\cite{konect:bibsonomy} & \textsf{\href{http://konect.uni-koblenz.de/networks/bibsonomy-2ui}{Bui}} & \href{http://konect.uni-koblenz.de/networks/bibsonomy-2ui}{BibSonomy user–publication} & B M & 777,084 & 2,555,080 \\
\cite{b349} & \textsf{\href{http://konect.uni-koblenz.de/networks/citeulike-ut}{Cut}} & \href{http://konect.uni-koblenz.de/networks/citeulike-ut}{CiteULike user–tag} & B M & 175,992 & 2,411,819 \\
\cite{b349} & \textsf{\href{http://konect.uni-koblenz.de/networks/citeulike-ui}{Cui}} & \href{http://konect.uni-koblenz.de/networks/citeulike-ui}{CiteULike user–publication} & B M & 754,484 & 2,411,819 \\
\bottomrule\toprule & \textcolor{colorAuthorship}{$\blacksquare$}  & \textbf{Wiki edit networks} & \textbf{Flags} & \textbf{$|V|$} & \textbf{$|E|$} \\ \midrule
\cite{download.wikimedia.org} & \textsf{\href{http://konect.uni-koblenz.de/networks/edit-frwikinews}{nfr}} & \href{http://konect.uni-koblenz.de/networks/edit-frwikinews}{Wikinews, French} & B M & 26,546 & 193,618 \\
\cite{download.wikimedia.org} & \textsf{\href{http://konect.uni-koblenz.de/networks/edit-frwikibooks}{bfr}} & \href{http://konect.uni-koblenz.de/networks/edit-frwikibooks}{Wikibooks, French} & B M & 30,997 & 201,727 \\
\cite{download.wikimedia.org} & \textsf{\href{http://konect.uni-koblenz.de/networks/edit-enwikiquote}{qen}} & \href{http://konect.uni-koblenz.de/networks/edit-enwikiquote}{Wikiquote, English} & B M & 116,363 & 549,210 \\
\cite{download.wikimedia.org} & \textsf{\href{http://konect.uni-koblenz.de/networks/edit-enwikinews}{nen}} & \href{http://konect.uni-koblenz.de/networks/edit-enwikinews}{Wikinews, English} & B M & 173,772 & 901,416 \\
\cite{download.wikimedia.org} & \textsf{\href{http://konect.uni-koblenz.de/networks/edit-dewiktionary}{mde}} & \href{http://konect.uni-koblenz.de/networks/edit-dewiktionary}{Wiktionary, German} & B M & 151,982 & 1,229,501 \\
\cite{download.wikimedia.org} & \textsf{\href{http://konect.uni-koblenz.de/networks/edit-enwikibooks}{ben}} & \href{http://konect.uni-koblenz.de/networks/edit-enwikibooks}{Wikibooks, English} & B M & 167,525 & 1,164,576 \\
\cite{download.wikimedia.org} & \textsf{\href{http://konect.uni-koblenz.de/networks/edit-frwiktionary}{mfr}} & \href{http://konect.uni-koblenz.de/networks/edit-frwiktionary}{Wiktionary, French} & B M & 1,912,264 & 7,399,298 \\
\cite{download.wikimedia.org} & \textsf{\href{http://konect.uni-koblenz.de/networks/edit-enwiktionary}{men}} & \href{http://konect.uni-koblenz.de/networks/edit-enwiktionary}{Wiktionary, English} & B M & 2,133,892 & 8,998,641 \\
\cite{download.wikimedia.org} & \textsf{\href{http://konect.uni-koblenz.de/networks/edit-itwiki}{it}} & \href{http://konect.uni-koblenz.de/networks/edit-itwiki}{Wikipedia, Italian} & B M & 2,393,568 & 26,241,217 \\
\cite{download.wikimedia.org} & \textsf{\href{http://konect.uni-koblenz.de/networks/edit-eswiki}{es}} & \href{http://konect.uni-koblenz.de/networks/edit-eswiki}{Wikipedia, Spanish} & B M & 3,288,398 & 27,011,506 \\
\cite{download.wikimedia.org} & \textsf{\href{http://konect.uni-koblenz.de/networks/edit-frwiki}{fr}} & \href{http://konect.uni-koblenz.de/networks/edit-frwiki}{Wikipedia, French} & B M & 4,310,551 & 46,168,355 \\
\cite{download.wikimedia.org} & \textsf{\href{http://konect.uni-koblenz.de/networks/edit-dewiki}{de}} & \href{http://konect.uni-koblenz.de/networks/edit-dewiki}{Wikipedia, German} & B M & 3,620,990 & 57,323,775 \\
\cite{download.wikimedia.org} & \textsf{\href{http://konect.uni-koblenz.de/networks/edit-enwiki}{en}} & \href{http://konect.uni-koblenz.de/networks/edit-enwiki}{Wikipedia, English} & B M & 25,323,882 & 266,769,613 \\
\bottomrule\toprule & \textcolor{colorContact}{$\blacksquare$}  & \textbf{Explicit interaction networks} & \textbf{Flags} & \textbf{$|V|$} & \textbf{$|E|$} \\ \midrule
\cite{b532} & \textsf{\href{http://konect.uni-koblenz.de/networks/contact}{HA}} & \href{http://konect.uni-koblenz.de/networks/contact}{Haggle} & U M & 274 & 28,244 \\
\cite{b665} & \textsf{\href{http://konect.uni-koblenz.de/networks/mit}{RM}} & \href{http://konect.uni-koblenz.de/networks/mit}{Reality Mining} & U M & 96 & 1,086,404 \\
\cite{b629} & \textsf{\href{http://konect.uni-koblenz.de/networks/munmun_twitterex_at}{Wa}} & \href{http://konect.uni-koblenz.de/networks/munmun_twitterex_at}{Twitter} & D M & 2,919,613 & 12,887,063 \\
\cite{b525} & \textsf{\href{http://konect.uni-koblenz.de/networks/dblp_coauthor}{Pc}} & \href{http://konect.uni-koblenz.de/networks/dblp_coauthor}{DBLP} & U M & 1,103,412 & 14,703,760 \\
\bottomrule\toprule & \textcolor{colorInteraction}{$\blacksquare$}  & \textbf{Implicit interaction networks} & \textbf{Flags} & \textbf{$|V|$} & \textbf{$|E|$} \\ \midrule
\cite{konect:opsahl10} & \textsf{\href{http://konect.uni-koblenz.de/networks/opsahl-ucforum}{UF}} & \href{http://konect.uni-koblenz.de/networks/opsahl-ucforum}{UC Irvine forum} & B M & 1,421 & 33,720 \\
\cite{said:social-similarity} & \textsf{\href{http://konect.uni-koblenz.de/networks/filmtipset_comment}{Fc}} & \href{http://konect.uni-koblenz.de/networks/filmtipset_comment}{Filmtipset} & B M & 75,360 & 1,266,753 \\
\cite{lastfm} & \textsf{\href{http://konect.uni-koblenz.de/networks/lastfm_band}{Lb}} & \href{http://konect.uni-koblenz.de/networks/lastfm_band}{Last.fm band} & B M & 175,069 & 19,150,868 \\
\cite{lastfm} & \textsf{\href{http://konect.uni-koblenz.de/networks/lastfm_song}{Ls}} & \href{http://konect.uni-koblenz.de/networks/lastfm_song}{Last.fm song} & B M & 1,085,612 & 19,150,868 \\
\bottomrule\end{tabular} }